\documentclass[aps,prl,twocolumn]{revtex4} 
\usepackage{amsmath} 
\usepackage{graphicx} 

\begin{document} 

\title{Effective Magnetic Monopoles and Universal Conductance Fluctuations} 

\author{Kjetil M.D. Hals$^{1}$, Anh Kiet Nguyen$^1$, Xavier Waintal$^2$ and Arne Brataas$^1$} 
\affiliation{$^1$ Department of Physics, Norwegian University of Science and  
Technology, NO-7491, Trondheim, Norway. \\
$^2$ SPSMS-INAC-CEA, 17 rue des Martyrs, 38054 Grenoble CEDEX 9, France} 
\begin{abstract}
The observation of isolated positive and negative charges, but not isolated magnetic north and south poles, is an old puzzle. Instead, evidence of effective magnetic monopoles has been found in the abstract momentum space. Apart from Hall-related effects, few observable consequences of these abstract monopoles are known. Here, we show that it is possible to manipulate the monopoles by external magnetic fields and probe them by universal conductance fluctuation (UCF) measurements in ferromagnets with strong spin-orbit coupling. The observed fluctuations are not noise, but reproducible quasiperiodic oscillations as a function of magnetisation direction, a novel Berry phase fingerprint of the magnetic monopoles.
\end{abstract}

\maketitle 

\newcommand{\eq}{\! = \!} 
\newcommand{\keq}{\!\! = \!\!} 
\newcommand{\kadd}{\! + \!} 
\newcommand{\ksim}{\! \sim \!} 

Quantum states in solids are classified by a crystal momentum vector and a band index. The space spanned by the momentum vectors is known as the momentum space. Each band index defines an energy-band of allowed electronic energy levels in the momentum space. Momentum-space magnetic monopoles arise from energy-band crossings ~\cite{Bohm:book03}. Each band crossing point produces a magnetic monopole with a quantised topological magnetic charge, characterised by a Chern number~\cite{Bohm:book03}. An electric particle traversing a closed curve in momentum space accumulates a geometric phase from the monopole fields~\cite{Berry:prca84,Sundaram:prb99}. So far, these abstract monopoles have revealed themselves only through Hall-related effects~\cite{Fang:science03, Nagaosa:RMF10}, but we show that they can also be manipulated and probed  by UCF measurements. 

UCFs are observed experimentally as reproducible fluctuations in the conductance in response to an applied external magnetic field $\rm{B}$~\cite{Lee:prb86}. The fluctuation pattern is known as the magneto-fingerprint of the sample~\cite{Lee:prb86}. Recent experiments on the ferromagnetic semiconductor (Ga, Mn)As report two different $\rm{B}_c$ periods in the conductance fluctuations~\cite{Vila:prl07,Neumaier:prl07}: a slow, conventional oscillation for high magnetic fields and a much faster oscillation for low fields when the magnetisation rotates. The present work reinterprets these recent experimental results and shows that the fast oscillations are caused by the relocation of momentum-space magnetic monopoles. Rotation of the magnetisation relocates the monopoles, which leads to a geometric phase change of the closed momentum-space curves. The numerical results demonstrate, in good agreement with the experiments, that a geometric phase change is observed with fast UCF oscillations, implying a novel Berry phase fingerprint of the monopoles.

The underlying physics of UCFs is quantum interference between different paths across the sample~\cite{Lee:prb86}.
Let $\rm{A}_c$ denote the quantum mechanical probability amplitude for propagating along the classical path $\mathbf{x}_c (t)$. The amplitude can be expressed as $\rm{A}_c= \left| \rm{A}_c\right| \exp(i\rm{S}[\mathbf{x}_c(t)]/\hbar)$ in terms of the action  $\rm{S}[\mathbf{x} (t)]= \int \rm{dt L }(\mathbf{x},\mathbf{\dot{x}})$, where $\rm{L}(\mathbf{x},\mathbf{\dot{x}})$ is the Lagrangian, $\mathbf{\dot{x}}\equiv \rm{d}\mathbf{x}/\rm{dt}$, and $\left| \rm{A}_c\right|^2$ is the probability to follow the path $\mathbf{x}_c(t)$~\cite{Rammer:book07}. When an external magnetic field $\mathbf{B}$ is applied, the term $-e\int \rm{dt}\mathbf{\dot{x}} \cdot \mathbf{A}$ should be added to the action~\cite{Rammer:book07}, where $e$ is minus the electron charge, and $\mathbf{A}$ is the vector potential corresponding to  $\mathbf{B}=\boldsymbol{ \nabla } \times\mathbf{A}$. Let us separate out the magnetic field-dependent phase and rewrite the amplitude as $\rm{A}_c = \tilde{\rm{A} }_c\exp (-i e /\hbar  \int \rm{dt}\mathbf{\dot{x}}_c\cdot \mathbf{A} )$. The conductance $\rm{G}$ is proportional to the total probability of propagating across the sample, $\rm{G(B)}\propto \left|\sum_c \rm{A}_c \right|^2$. Reformulating the line integral associated with the vector potential as a surface integral using Green's theorem, one finds $\rm{G(B)}\propto   \sum_{c c^{'}} \tilde{\rm{A}}_c^{*} \tilde{\rm{A}}_{c^{'}} \exp(i 2\pi \Phi_{c c^{'}}(B)/\Phi_0  ) $, where $\Phi_{c c^{'}}(B)$ is the magnetic flux enclosed by the loop formed by the paths $\mathbf{x}_{c}(t)$ and $\mathbf{x}_{c^{'}}(t)$ and $\Phi_0 \equiv h/e$. Changing the magnetic field randomises the phase difference between different pairs of paths, causing the conductance to fluctuate. A typical period $\rm{B}_c$ of these quasiperiodic oscillations is when the dominant paths experience a relative phase shift of $2\pi$. Assuming that typical paths approximately enclose the sample area $\mathcal{A}$, leads to $\rm{B}_c = \Phi_0 / \mathcal{A}$~\cite{Lee:prb86}.      

A closed loop in real space also corresponds to a closed loop in momentum space. 
In systems with either  broken inversion or time-reversal symmetry, there is also a phase associated with paths in momentum space~\cite{Sundaram:prb99,Bohm:book03}. Semiclassically, this Berry phase effect is included in the Lagrangian as $\hbar\mathbf{A}^{(n)}\cdot \mathbf{\dot{k}}$, where  $\mathbf{A}^{(n)} (\mathbf{k})=i \left\langle u_n  \left| \boldsymbol{ \nabla }_{\mathbf{k}} \right| u_n \right\rangle$ is the Berry connection, $\left| u_n \right\rangle$ is the periodic part of the Bloch function, and $n$ is the band index~\cite{Sundaram:prb99}. The propagation amplitudes accumulate a geometric phase factor  $\exp(i\int \rm{d}\mathbf{k}\cdot \mathbf{A}^{(n)}(\mathbf{k}))$ along a path in momentum space. A closed momentum-space curve acquires a phase equal to the flux of the effective field $\mathbf{\Omega}^{(n)} (\mathbf{k})= \boldsymbol{ \nabla }_{\mathbf{k}}\times \mathbf{A}^{(n)}(\mathbf{k})$ that the loop encloses~\cite{Sundaram:prb99,Bohm:book03}. The effective field is known as the Berry curvature~\cite{Berry:prca84,Bohm:book03}: 
\begin{eqnarray} 
\mathbf{\Omega}^{(n)} (\mathbf{k}) & = & i\sum_{m\neq n}  
\frac{  \left\langle u_n  \left|  \boldsymbol{ \nabla }_{\mathbf{k}} H    \right|   u_m \right\rangle \times  
\left\langle u_m  \left|  \boldsymbol{ \nabla }_{\mathbf{k}} H    \right|   u_n \right\rangle    }{ \left( \rm{E}_{n}(\mathbf{k}) - \rm{E}_{m}(\mathbf{k}) \right) ^2 },	\label{BerryCurvature}		   
\end{eqnarray}	 
where $H$ is the Hamiltonian of the system and $\rm{E}_n(\mathbf{k})$ is the dispersion relation of the $n$th band. Momentum-space magnetic monopoles are singularities in the Berry curvature where energy bands cross at isolated points~\cite{Fang:science03,Nagaosa:RMF10}. In ferromagnets with strong spin-orbit coupling, the Hamiltonian is not invariant under rotation of the magnetisation~\cite{Jungwirth:RMP06}. Changing the magnetisation direction relocates the magnetic monopoles, inducing a geometric phase change in the propagation amplitudes. The external magnetic field can rotate the magnetisation in UCF experiments on ferromagnets. The phase change of a closed real-space curve then also acquires important contributions from the geometric phase change of the corresponding closed momentum-space curve. 
We demonstrate that the magnetic monopoles give rise to fast conductance oscillations at low magnetic fields. This novel and large magnetic monopole
effect is qualitatively different from the studies of Berry phase effects in two-dimensional electron gases with Rashba spin-orbit coupling since these
systems exhibit no effective momentum-space monopoles~\cite{Engel:prb00}. Also, the effect we compute quantitively differs from the weak peak splitting effects seen
therein by 1 order of magnitude.

In the following discussion, the Berry phase effect on UCFs will be investigated for the ferromagnetic semiconductor (Ga, Mn)As. The system is modeled by the Hamiltonian~\cite{Jungwirth:RMP06} 
\begin{equation} 
 H = (\gamma_1 + \frac{5}{2} \gamma_2) \frac{\mathbf{P}^2}{2 m_e}  
 - \frac{\gamma_2}{m_e} (\mathbf{P} \cdot \mathbf{J})^2  
 + \mathbf{h} \cdot \mathbf{J} + V(\mathbf{r}). 
\label{Hamiltonian} 
\end{equation} 
The band structure of the host compound is described by the two first terms in Eq.~\eqref{Hamiltonian}, characterised by the Luttinger parameters $\gamma_1$ and $\gamma_2$. $\mathbf{J}$ is a vector of $4 \!  \times \! 4$ spin matrices for 3/2 spins, and $\mathbf{P}= \mathbf{p} - e \mathbf{A}$ is the canonical momentum operator in the presence of an external magnetic field $\mathbf{B}=\boldsymbol{ \nabla } \times\mathbf{A}$. $m_e$ denotes the electron mass. The third term describes the exchange interaction between the holes and the local magnetic moments, modeled by a homogenous exchange field $\mathbf{h}$. To model disorder, we used the impurity potential $V(\mathbf{r}) = \sum_{i} V_i  \delta(\mathbf{r}-\mathbf{R}_i)$, where $V_i$ and $\mathbf{R}_i$ are the strength and position of impurity number $i$ and $\delta(\mathbf{r})$ is the delta 
function. 

The magnetocrystalline anisotropy in (Ga, Mn)As is complicated and depends on several material parameters such as doping, strain and shape: see Ref. \cite{Jungwirth:RMP06} and references therein. We consider two cases: 1) a perpendicular easy magnetisation axis that is valid, for example, for (Ga, Mn)As
grown on (Ga, In)As and 2) a uniaxial in-plane easy
magnetisation axis that is valid for (Ga, Mn)As bars grown on a GaAs
substrate~\cite{Jungwirth:RMP06}. Here, the magnetisation and hence $\mathbf{h}$
are assumed to be governed by the following  magnetic free energy $\varepsilon = K_u \sin^2(\phi_M) - M B \cos(\phi_M - \phi_B)$, where $\phi_M$ ($\phi_B$) is the angle between the exchange field $\mathbf{h}$ (applied magnetic field $\mathbf{B}$) and
the current.

For the numerical UCF calculation, we considered a discrete rectangular conductor 
sandwiched between two clean reservoirs with rectangular cross sections defined by $L_x = 30~nm$ and $L_z = 14~nm$. 
The spacing between the lattice points is $a_x = a_y = a_z = 
1~nm$, significantly smaller than the typical Fermi wavelengths used at
$\lambda_F \sim 5~nm$. We assumed one impurity at each lattice site 
in the conductor. The current direction is $[010]$, and the crystal growth direction and applied magnetic field are along $[001]$.  We used the Landau gauge $\mathbf{A} = B x \hat{y}$. For direct comparisons with experimental findings, we used parameters 
appropriate for (Ga, Mn)As: $\gamma_1 = 
7.0$, $\gamma_2 = 2.5$, Fermi energy $E_F = 78~meV$ and $|\mathbf{h}| = 31~meV$.  
The impurity strengths $V_i$ are uniformly distributed between $-V_0/2$ and $V_0/2$. $V_0 = 0.75~eV$, which leads to a mean free 
path of $\sim 6~nm$. We assume that $M = 2 \times 10^{4}~A/m$ 
and the uniaxial anisotropy constant $K_u = 5 \times 10^{3}~J/m^3$, 
giving the anisotropy field $B_u=2 K_u / M = 0.5~T$, which is similar to the experimental 
value found in Ref.~\cite{Vila:prl07}.  
The Landauer-B{\" u}tikker formula is used to calculate  
the conductance from a stable transfer matrix method~\cite{Usuki:prb95}.
More details about the numerical calculation method can be found in Ref.~\cite{Nguyen:prl08}.

\begin{figure}[ht] 
\centering 
\includegraphics[scale=0.9]{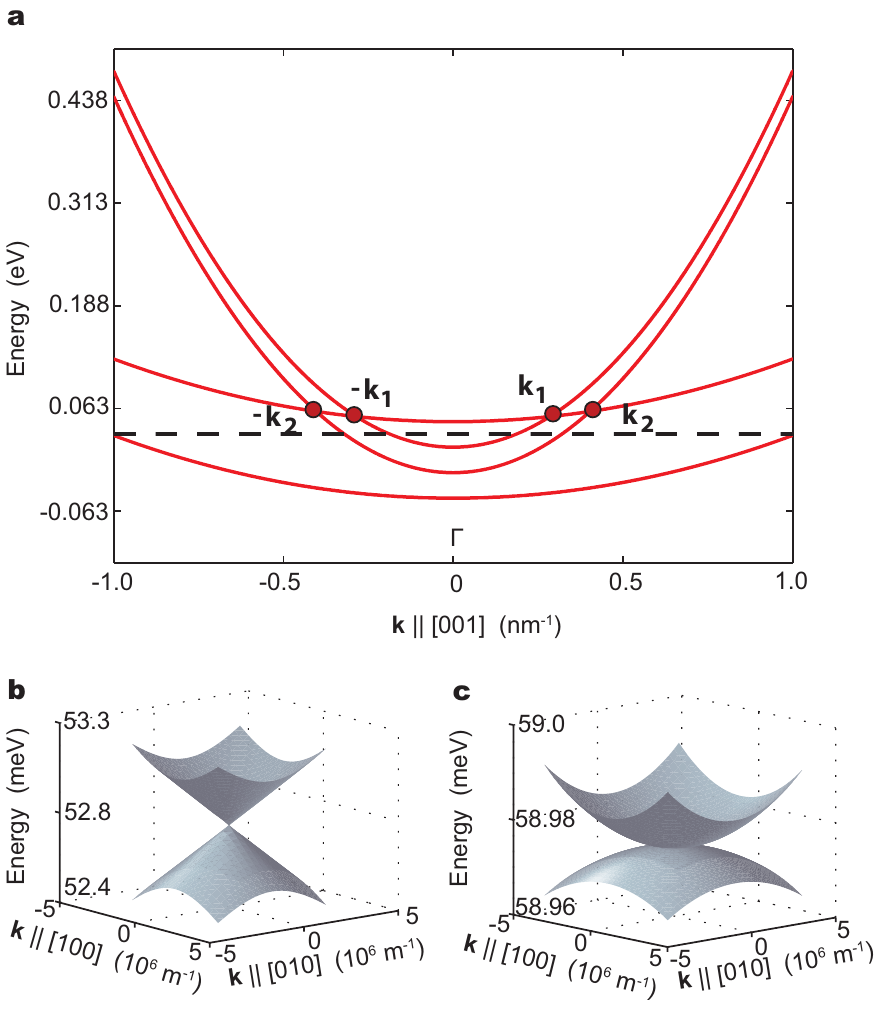}  
\caption{ (\textbf{a}) Dispersion curves along the $[001]$ axis when the exchange field $\mathbf{h}$ is pointing along $[001]$. Each band crossing point $\pm \mathbf{k}_{1,2}$ gives rise to a momentum-space monopole. The black dotted line is the Fermi level used in the numerical UCF simulation. (\textbf{b}) Energy surfaces in the $k_x k_y$ plane near the $\mathbf{k}_1$ band crossing point in figure (a). (\textbf{c}) Energy surfaces in the $k_x k_y$ plane near the $\mathbf{k}_2$ band crossing point in figure (a).}
\label{Fig1} 
\end{figure} 
Let us first analyse and classify the monopoles and then use simple semiclassical considerations to estimate the Berry phase-induced UCF oscillation period. The geometric phase-induced conductance oscillations appear for weak external magnetic fields, and we can therefore neglect the real-space magnetic field in the analysis of the Berry curvature. Without magnetic fields and disorder, the Hamiltonian in Eq.~\eqref{Hamiltonian} has four band crossing points located at $\pm | \mathbf{k}_1 | = \pm \sqrt{ ( \left| \mathbf{h} \right| m_e ) / ( 2\gamma_2 \hbar^2 )}$ and $\pm | \mathbf{k}_2 | = \pm \sqrt{ ( \left| \mathbf{h} \right| m_e ) /  (\gamma_2 \hbar^2 )}$ along the momentum-space axis parallel to the exchange field, as shown in Fig.~\ref{Fig1}a. Each crossing point gives rise to a magnetic monopole. The eigenfunctions of the Hamiltonian are of the form $\exp(i\mathbf{k\cdot r})\, \boldsymbol{\chi}_{n,\mathbf{k}}$ where $\boldsymbol{\chi}_{n,\mathbf{k}}$ is a four-component spinor. Because the helicity operator $\hat{\Sigma} \equiv \mathbf{k}\cdot \mathbf{J} / \left| \mathbf{k} \right|$ commutes with the Hamiltonian when $\mathbf{k}$ is parallel to $\mathbf{h}$,  the eigenspinors along this axis in momentum space are $\boldsymbol{\chi}_{m_z, \mathbf{k || h }} = \hat{U}(\theta,\phi) \left| m_z  \right\rangle$ where $\left| m_z  \right\rangle$  are eigenvectors of $\hat{J}_z$ and  $\hat{U}(\theta,\phi)$ is the unitary rotation operator that rotates the quantisation axis parallel to $\mathbf{h}$. At the point $\mathbf{k}_1$, $\boldsymbol{\chi}_{3/2}$ and $\boldsymbol{\chi}_{1/2}$ are degenerate eigenspinors. Close to this point, the Hamiltonian couples these two states only weakly to the $\boldsymbol{\chi}_{-3/2}$ and $\boldsymbol{\chi}_{-1/2}$ states, and it can therefore be written as a $2\times 2$ matrix in the basis of $ \left( 1 \  0 \right)^T\equiv \boldsymbol{\chi}_{1/2} $ and $  \left( 0 \  1 \right)^T \equiv \boldsymbol{\chi}_{3/2}  $. Expanding the Hamiltonian around the degenerate point, $H= H(\mathbf{k}_1) + \delta \mathbf{k}\cdot \boldsymbol{\nabla}_{\mathbf{k}_1} H$ ($\delta \mathbf{k} \equiv \mathbf{k} - \mathbf{k}_1$), treating the last term as a perturbation and considering the case when $\mathbf{h} = \left| \mathbf{h} \right| \hat{\mathbf{z}}$, we obtain the local $2\times 2$ Hamiltonian:
\begin{equation} 
 \rm{H} = \rm{E_0} \left( \delta  \mathbf{k} \right)\hat{I}   + \frac{1}{2}\mathbf{x}\cdot\boldsymbol{\sigma}, 
\label{LocalHamiltonian} 
\end{equation} 
where $\rm{E_0} \left(   \delta \mathbf{k} \right)$ is an energy shift of the two energy bands away from the source point, $\boldsymbol{\sigma}$ is a vector of Pauli matrices, $\rm{\hat{I}}$ is the identity matrix, and $\mathbf{x}\equiv 2\gamma_2 \hbar^2 \left| \mathbf{k}_1 \right| / m_e \left( -\sqrt{3} \delta k_x,\:   \sqrt{3} \delta k_y, \:  2\delta k_z   \right)$. 
When the crystal momentum $\mathbf{k}$ varies in time, the effective Hamiltonian in Eq.~\eqref{LocalHamiltonian} describes a spin in a time-varying magnetic field and the electron accumulates a well-known geometric phase from the Berry curvature field~\cite{Berry:prca84}  
\begin{equation} 
 \boldsymbol{\Omega}^{( \pm )} \left( \mathbf{k} \right) = \mp \frac{\mathbf{k} }{ 2 \left| \mathbf{k} \right|^3} , 
\label{} 
\end{equation} 
where $\mp$ refer to the upper and lower energy bands near $\mathbf{k}_1$. We have here reparameterised the momentum space as $\mathbf{k}\mapsto \mathbf{k}=  \left( -x_1,\: x_2,\: x_3   \right)$ for clarity.  
The topological magnetic charge of this monopole, its Chern number, is $1/2\pi \int_{S^2} \boldsymbol{\Omega}^{( \pm )}\cdot\mathbf{\hat{n}}\, \rm{dS} = \mp 1$~\cite{Bohm:book03}. 

Because the Berry curvature is inversely proportional to $\left( E_{n}(\mathbf{k}) - E_{m}(\mathbf{k}) \right)^2$, 
bands that nearly cross over a larger region in momentum space produce a stronger monopole. 
The structures of the energy bands in Figs.~\ref{Fig1}b-c show that the $\mathbf{k}_2$ monopole is stronger than the $\mathbf{k}_1$ monopole. A similar simple perturbative analysis of the $\mathbf{k}_2$ monopole cannot be carried out, but numerically, we find that the curvature decays asymptotically as $\mathbf{k}^{-2}$, and have a Chern number of $\pm 2$. 

 
As can be seen from Fig.~\ref{Fig1}a, the Berry curvature field from the two $\pm \mathbf{k}_1$ monopoles is experienced by orbits in the third and fourth bands, whereas the  $\pm \mathbf{k}_2$ monopoles give a geometric phase effect only to momentum-space curves in the second and third bands. In the third band, the $\pm \mathbf{k}_1$ and $\pm \mathbf{k}_2$ monopoles have topological charges of opposite signs and therefore counteract each other. Orbits in the lowest band do not experience any monopole field. Therefore, paths on the Fermi surface of the second band, which are experiencing the strong $\pm \mathbf{k}_2$ monopoles, dominate the Berry phase-induced conductance fluctuations.

\begin{figure}[ht] 
\centering 
\includegraphics[scale=0.85]{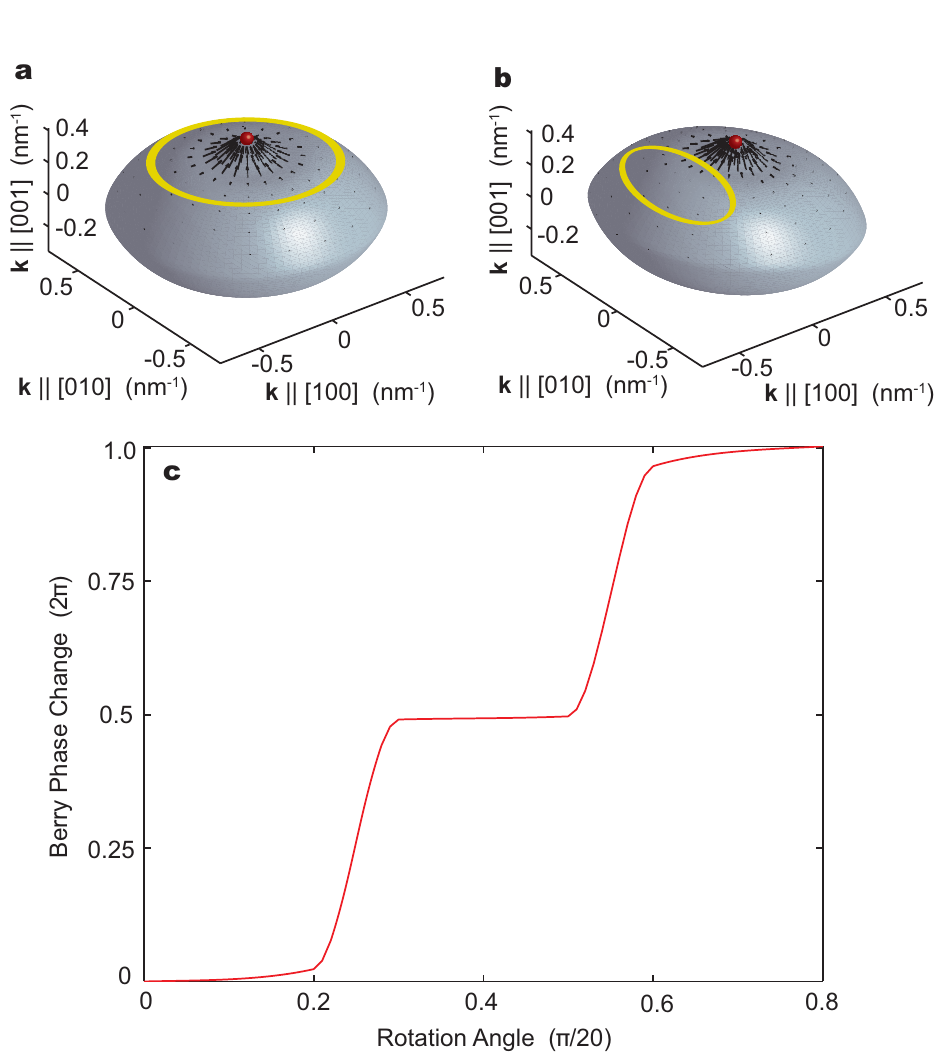} 
\caption{(\textbf{a}) The second-band Fermi surface and the Berry curvature field (black arrows) on this surface when the exchange field points along $[001]$. A typical path is shown in (a) as a yellow curve where $k_z \sim \pi/ L_z$. The curve accumulates a geometric phase of $\sim 2\pi$. (\textbf{b}) The second-band Fermi surface and the Berry curvature field (black arrows) on this surface when the exchange field is rotated by an angle of $0.13$. The yellow curve represents the same real-space curve as in (a). It accumulates a vanishing geometric phase. In both plots, the red dot is the $\mathbf{k}_2$ monopole source. (\textbf{c}) The geometric phase change of the curve as a function of the rotation angle of the exchange field. The geometric phase angle and rotation angle are in units of, respectively, $2\pi$ and $\pi / 20$ rad.}
\label{Fig2} 
\end{figure}
A typical closed momentum-space curve on the second-band Fermi surface is shown in Fig.~\ref{Fig2}a. The corresponding real-space curve is found from the semiclassical equation $\hbar\mathbf{\dot{r}}=\boldsymbol{\nabla}_{\mathbf{k}}\rm{E}_2(\mathbf{k})|_{\rm{E_2=E_F}}$. Because the Fermi surface is not rotationally symmetric, the momentum-space curve corresponding to this fixed real-space curve changes location on the Fermi surface when $\mathbf{h}$ is rotated, as illustrated in Fig.~\ref{Fig2}b. The associated geometric phase change of the relocated momentum-space curve, calculated numerically, is shown in  Fig.~\ref{Fig2}c. We found that a rotation angle on the order of $0.13$ changes the Berry phase of this fixed real-space curve by $2\pi$. The origin of the rapid phase change occurs when the curve in Fig.~\ref{Fig2}a encloses the strong Berry curvature field region on the Fermi surface, shown as black arrows in Fig.~\ref{Fig2}a-b, whereas the curve in Fig.~\ref{Fig2}b is relocated outside this region.

The magnetic field needed to rotate the magnetisation by $0.13$ rad is therefore an estimate of the oscillation period of the Berry phase fingerprint. Using the free energy defined above, this leads to the oscillation period  $B_c^{\rm{Berry}} \sim B_u /10$, where $B_u$ is the minimal external magnetic field needed to align the magnetisation along the hard magnetic anisotropy axis.

\begin{figure}[ht] 
\centering 
\includegraphics[scale=0.9]{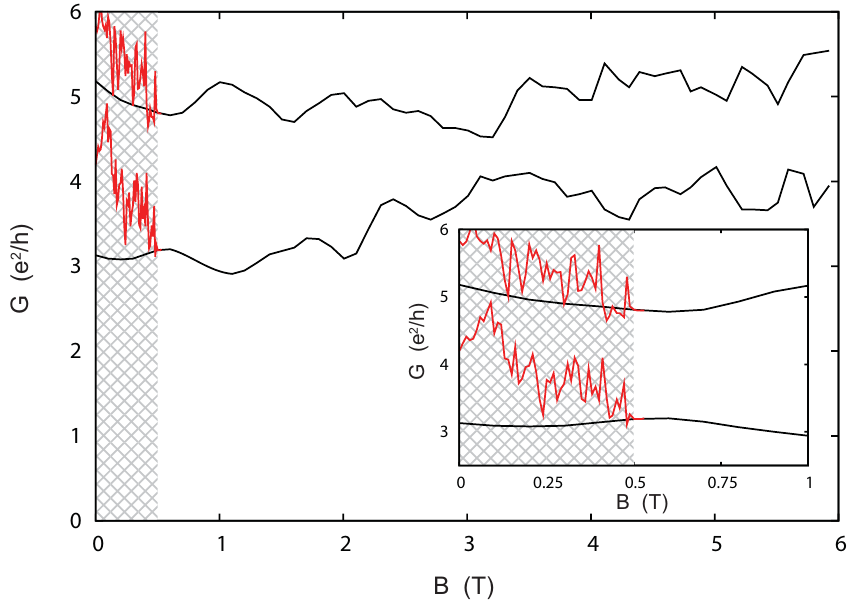} 
\caption{The conductance versus the applied magnetic field for the perpendicular magnetisation easy axis (black) and in-plane easy axis (red) for two different impurity configurations. The upper curve is shifted $1.5 e^2/h$ upwards for clarity. The grey region marks the region where the sample's magnetisation changes from in-plane to perpendicular. The curves for the perpendicular and in-plane easy axes are coincident for $B > B_u = 0.5T $ because the same impurity configuration is used. Inset: a close-up for small magnetic fields.}
\label{fig:Ghy} 
\end{figure}  
Let us next confirm our semiclassical analysis by a numerical UCF simulation of the system including disorder and magnetic field in Eq.~\eqref{Hamiltonian}.
We consider two cases: 1) a perpendicular easy magnetisation axis that is valid, for example, for (Ga, Mn)As
grown on (Ga, In)As and 2) a uniaxial in-plane easy
magnetisation axis that is valid for (Ga, Mn)As bars grown on a GaAs
substrate~\cite{Jungwirth:RMP06}. The external magnetic field is applied along the $[ 0 0 1]$ growth direction.

First, consider the case of a perpendicular magnetisation easy
axis where the magnetisation is aligned along the growth direction.
We see in Fig.~\ref{fig:Ghy} that the conductance has a weak increasing trend for
increasing $B$. Here, the magnetic field squeezes the spatial
extension of the wave function~\cite{Datta:book95}, allowing more conducting channels to be
open for increasing $B$.  Imposed on the increasing trend, there are
strong conductance fluctuations with a dominant period $B_c \sim
1T$. Because the magnetisation here is always along $[0 0 1]$, the
only change in the quantal phases comes from the magnetic flux.  For a wire of $\mathcal{A} = 30
\times 100nm^2$ $xy$ area, the dominant fluctuation period is $\rm{B}_c = \Phi_0 / \mathcal{A}\sim 1T$~\cite{Lee:prb86}, consistent with the data shown in Fig.~\ref{fig:Ghy}. 

Second, consider the case of an in-plane easy axis where $\mathbf{h}$ rotates from the $[0 1 0]$ direction to the hard $[ 0 0 1]$ axis when the magnetic field increases from $0$ to the anisotropy field $B_u = 0.5T$. The decreasing
trend of the conductance for increasing $B \in [0,B_u]$, shown in Fig. \ref{fig:Ghy}, is the standard anisotropic magnetoresistance effect~\cite{Nguyen:prl08}. Imposed on the decreasing trend,
the conductance fluctuates wildly for $B < B_u$ with a period on the order of $B_u /10$. Here, changing $B$ leads to changes in the direction of $\mathbf{h}$, which {\em relocates the position of the momentum-space magnetic monopoles and thereby the geometric phase for a given real-space orbit}. This gives rise to the extraordinarily fast conductance fluctuations shown in Fig. \ref{fig:Ghy}. For $B > B_u$, the Berry phase is fixed and the UCF again exclusively comes from the conventional magnetic flux.
 
Similar to what is found for the intrinsic anomalous Hall effect~\cite{Fang:science03, Nagaosa:RMF10}, the effect is strongest for Fermi energies near the monopole sources. We expect the effect to also be present for more highly doped (Ga, Mn)As systems that require a six- or eight-band model in which more monopoles are expected to exist. 
 
In conclusion, the UCF simulation in Fig.~\ref{fig:Ghy} semiquantitatively reproduces the experiments in Refs.~\cite{Vila:prl07,Neumaier:prl07}, and together with our semiclassical analysis, it reinterprets the fast oscillations as a Berry phase fingerprint.

This work was supported by computing time through the Notur project.

\end{document}